\documentclass[review,12pt]{elsarticle}
\usepackage{lineno,hyperref}
\usepackage{graphicx}
\usepackage{geometry}
\usepackage{caption}
\usepackage{subcaption}
\usepackage[linesnumbered,ruled,lined]{algorithm2e}
\usepackage{float}
\usepackage{tikz}
\usepackage{epstopdf}
\usepackage{amssymb}
\usepackage{amsmath}
\usepackage{indentfirst}
\usepackage{subcaption}
\usepackage{footnote}
\usepackage{array}
\usepackage{cases}
\usepackage{multirow}
\usepackage{ upgreek }
\usepackage{makecell}
\begin{document}
%
%
\begin{frontmatter}
    \title{Incorporating the algorithm for the boundary condition from FVM into the framework of Eulerian SPH}
    
    \author{Zhentong Wang }
    \ead{zhentong.wang@tum.de}
    \author{Oskar J. Haidn}
    \ead{oskar.haidn@tum.de}
    \author{Xiangyu Hu \corref{mycorrespondingauthor}}
    \ead{xiangyu.hu@tum.de}
    \address{TUM School of Engineering and Design, 
    Technical University of Munich, Garching, 85747, Germany}
    \cortext[mycorrespondingauthor]{Corresponding author. }
    
    \begin{abstract}
    Finite volume method (FVM) is a widely used mesh-based technique, renowned for its computational efficiency and accuracy   
    but it bears significant drawbacks, 
    particularly in mesh generation and handling complex boundary interfaces or conditions. 
    On the other hand, 
    smoothed particle hydrodynamics (SPH) method, a popular meshless alternative, 
    inherently circumvents the mesh generation and yields smoother numerical outcomes but at the expense of computational efficiency. 
    Therefore, 
    numerous researchers have strategically amalgamated the strengths of both methods to investigate complex flow phenomena 
    and this synergy has yielded precise and computationally efficient outcomes. 
    However,
    algorithms involving the weak coupling of these two methods tend to be intricate, 
    which has issues pertaining to versatility, implementation, and mutual adaptation to hardware and coding structures.
    Thus, 
    achieving a robust and strong coupling of FVM and SPH in a unified framework is imperative. 
    Wang \cite{wang2023extended} has integrated mesh-based FVM into the SPH library SPHinXsys.  
    However, 
    due to differing boundary algorithms between these methods in Wang's work, 
    the crucial step for establishing a strong coupling of both methods within a unified SPH framework 
    lies in incorporating the FVM boundary algorithm into the Eulerian SPH method.
    In this paper, 
    we propose a straightforward algorithm in the Eulerian SPH method, 
    algorithmically equivalent to that in FVM, 
    grounded in the principle of zero-order consistency. 
    Moreover, 
    several numerical examples, 
    including fully and weakly compressible flows with various boundary conditions in the Eulerian SPH method, 
    validate the stability and accuracy of the proposed algorithm.
    \end{abstract}
    
    \begin{keyword}
    Boundary algorithm  \sep Finite volume method \sep Smoothed particle hydrodynamics \sep Strong coupling 
    \sep Zero-order consistency \sep SPHinXsys
    \end{keyword}
    \end{frontmatter}
%
%
\section{Introduction}\label{referebce}
In the context of rapidly advancing high-performance computing, 
computational fluid dynamics (CFD) has emerged as a highly promising method for addressing industrial challenges 
and has proven instrumental in elucidating longstanding flow phenomena across a spectrum of scales, 
ranging from microfluidics to hydrodynamics and hypersonics \cite{afshari2018numerical,o2022eulerian}. 
While the traditional mesh-based CFD methods have undoubtedly achieved significant accomplishments, 
they exhibit drawbacks when confronted with complicated problems in practice. 
These challenges encompass the generation of high-quality meshes 
and the constraints arising from modelling complicated interfaces 
such as solid-fluid moving boundary \cite{mahady2015volume} and violent deformations (e.g. wave breaking).
Fortunately, 
for offering a solution to the latter constraints, 
the traditional mesh-based techniques, such as finite volume method (FVM), 
can be coupled with other numerical techniques, 
including mesh-based \cite{kim2001immersed,roman2009improved,erath2013posteriori,cheng2022numerical} 
and meshless \cite{li2019coupled,chiron2018coupled,napoli2016coupled} methods and then enable obtaining convincing and accurate results.
However, 
the former challenge is inherent in complex geometric settings within mesh-based methods. 
With the properties of avoiding the mesh generation and allowing large deformations during the simulations, 
the meshless methods have garnered significant interest 
due to its numerical formulation being rooted in particles and not contingent on the topology defined by a mesh structure. 
As one of the typical meshless methods, 
smoothed particle hydrodynamics (SPH) method is characterized by numerical approximations 
grounded in Gaussian-like kernel functions \cite{gingold1977smoothed,lucy1977numerical} 
and has found widespread application in diverse fields including CFD \cite{monaghan1994simulating}, structural mechanics \cite{libersky2005smooth}, 
and various scientific and engineering domains \cite{zhang2022review,gotoh2021entirely}, 
particularly in scenarios where traditional mesh-based methods encounter challenges.

As previously indicated, 
numerous researchers have integrated FVM with meshless approaches to tackle intricate flows characterized by complex interfaces or boundaries. 
Similarly, 
they also have combined SPH with traditional mesh-based methods to enhance computational efficiency and address intricate boundary conditions, 
as documented in the literature \cite{chiron2018coupled,napoli2016coupled,wang2021graphics}. 
Consequently, 
the fusion of mesh-based methods such as FVM and meshless methods such as SPH is of paramount significance for solving complex fluid dynamics problems. 
Nevertheless, it is essential to acknowledge that weakly coupling these multiple methods also introduces inherent limitations.
Firstly, 
a potential concern pertains to the decreased numerical stability inherent in the amalgamation of two distinct methods 
when compared to the application of a single method, 
which arises from the necessity to formulate tailored algorithms 
for the purpose of data transfer or exchange within the overlapping regions or interfaces \cite{li2019coupled,napoli2016coupled,li2023coupled,marrone2016coupling}.
Secondly, 
the implementation and generalization of coupling algorithms are intricate tasks \cite{vacondio2021grand}, 
primarily owing to the inherent divergence in the fundamental characteristics of the coupled models.
Thirdly, 
apart from variations in formulations, 
another notable challenge arises when attempting to couple methodologies 
that are tailored or extensively optimized for distinct types of hardware acceleration and coding constructs \cite{vacondio2021grand}.
Hence, 
to address these intricate limitations,  
introducing a single framework for computing complicated CFD problems emerges as an imperative challenge that requires resolution.
Following Ref. \cite{wang2023extended},
Wang has proposed a certain form of Eulerian SPH method equivalent to FVM and implemented mesh-based FVM within an open-source SPH library (SPHinXsys), 
which paves the path of strong coupling these two methods within a unified SPH framework.
While the both methods in the article \cite{wang2023extended} share the same discretized conservation equations 
and other techniques for improving the stability and accuracy, 
it is important to note that the treatment of boundary conditions varies between the particle-based and mesh-based approaches. 
Based on this, 
the incorporation of FVM boundary algorithm into Eulerian SPH method constitutes a pressing concern demanding immediate attention, 
as it not only enhances the equivalence between the Eulerian SPH and FVM 
but also constitutes a crucial component in establishing a robust and unified SPH framework for achieving a stronge coupling between these two methods.

In this paper, 
by analysing the boundary algorithm in FVM where the each cell in the boundaries obeys the zero-order consistency, 
we propose a simple and general boundary algorithm adapted to Eulerian SPH method. 
During the process of implementing the boundary algorithm in FVM within the SPHinXsys,
we create the extra storage and save the necessary data to keep the zero-order consistency 
for the particles within the boundaries missing several neighbouring particles. 
Besides, 
several numerical examples with a series of boundary conditions, including the reflecting, inflow, 
outflow, motiontive, non-reflecting, non-slip wall and far-field boundary conditions, 
are tested in Eulerian SPH method to validate the stability and accuracy of the proposed algorithm.

The article is organized as follows: the governing equation and its discretization are given.
Also,
the dissipation limiters for improved accuracy and particle relaxation for the body-fitted particle distribution 
and zero-order consistency are presented in Chapter \ref{Governing equations and SPH method}. 
Besides, 
Chapter \ref{Eulerian general boundary conditions} explains the process of implementing the new boundary algorithm in Eulerian SPH in detail 
by comparing that in FVM. 
Finally, 
several examples with different boundary conditions are tested for the performance of the proposed algorithm in Chapter \ref{Numerical results}.
The computational codes used in this study are readily available through the SPHinXsys repository \cite{zhang2021sphinxsys,zhang2020sphinxsys}, 
which can be accessed via the following URLs: https://www.sphinxsys.org and https://github.com/Xiangyu-Hu/SPHinXsys.
%
%
\section{Governing equations and Eulerian SPH method}\label{Governing equations and SPH method}
%
%
\subsection{Governing equations}\label{Governing equations}
The conservation equations within the Eulerian framework can be delineated as follows:
\begin{equation}\label{eqs:conservation}
\frac{\partial \boldsymbol{U}}{\partial t}+\nabla \cdot \boldsymbol F(\boldsymbol{U})=0,
\end{equation}
where $\boldsymbol{F}(\boldsymbol{U})$ is the corresponding fluxes of $\boldsymbol{U}$ denoting the vector of the conserved variables.
In a two dimensional situation, 
they are expressed as 
\begin{equation}
\label{eqs:flux}
\mathbf{U}=\left[\begin{array}{c}
\rho \\
\rho u \\
\rho v \\
E
\end{array}\right], \quad \mathbf{F}=\left[\begin{array}{c}
\rho u \\
\rho u^{2}+p \\
\rho u v \\
u(E+p)
\end{array}\right]+\left[\begin{array}{c}
\rho v \\
\rho vu \\
\rho v^{2}+p \\
v(E+p)
\end{array}\right], 
\end{equation}
where $u$ and $v$ are the components of velocity, $\rho$ and $p$ denote the density and pressure, respectively, 
as well as $E=\frac{\rho{\mathbf{v}}^2}{2}+\rho e$ is the total energy with $e$ the internal energy.
The equation of state (EOS) is introduced to close Eq. \eqref{eqs:conservation} as follows:
\begin{equation}
p=\begin{cases} \rho(\gamma-1)e & \text { For compressible flows } \\ c^2(\rho-\rho_0) & \text { For weakly-compressible flows } \end{cases}. 
\end{equation}
Here, $\gamma$ represents the heat capacity ratio, and $\rho_{0}$ is the reference density.
For compressible flows, we follow the ideal gas equation to derive the speed of sound:
\begin{equation}\label{sound equation}
    c^2=\frac{\gamma p}{\rho}.
\end{equation}
In the case of incompressible flows, 
we adopt the weakly compressible assumption by employing an artificial EOS: 
\begin{equation}
p=c^2(\rho-\rho_0). 
\end{equation}
To restrict density variations to within 1\%, 
we define the speed of sound $c=10U_{max}$, where $U_{max}$ signifies the anticipated maximum velocity within the flow field. 
It is noteworthy that in this weakly compressible formulation, 
we turn off the energy conservation equation in Eq. \eqref{eqs:flux}.
%
%
\subsection{Riemann-based Eulerian SPH discretization}\label{Standard Eulerian SPH discretization}
In accordance with Ref. \cite{vila1999particle}, we can express the discretized form of Eq. \eqref{eqs:conservation} as follows:
\begin{equation}\label{eqs:conservation-discretize}
\left\{\begin{array}{l}
\frac{\partial}{\partial t}\left(w_{i}\rho_{i}\right)+2 w_{i}\sum_{j} w_{j}  (\rho \mathbf{v})^{*}_{E, i j} \cdot \nabla W_{i j}=0 \\
\frac{\partial}{\partial t}\left(w_{i}\rho_{i} \mathbf{v}_{i}\right)+
2 w_{i}\sum_{j} w_{j} \left[(\rho \mathbf{v} \otimes \mathbf{v})^{*}_{E, i j}+p^{*}_{E, i j}\mathbb{I}\right] \cdot \nabla W_{i j}=0 \\
\frac{\partial}{\partial t}\left(w_{i}E_{i}\right)+ 2 w_{i}\sum_{j} w_{j} \left[(E\mathbf{v})^{*}_{E, i j}+
(p \mathbf{v})^{*}_{E, i j}\right] \cdot \nabla W_{i j}=0
\end{array}.\right.
\end{equation}
In this formulation, 
$w$ signifies the particle volume, $\mathbf{v}$ denotes the velocity, $\mathbb{I}$ stands for the identity matrix, 
and $\nabla W_{i j}=\frac{\partial W_{i j}}{\partial r_{ij}}\mathbf{e}_{ij}$ is the gradient of the kernel, 
where $\mathbf{e}_{ij}=-\mathbf{r}_{ij}/r_{ij}$ and $\mathbf{r}_{ij}$ represents the displacement vector pointing from particle $j$ to $i$.
Moreover, the terms $()^{*}_{E, i j}$ correspond to the solutions of the Riemann problem \cite{vila1999particle}.

To solve the Riemann problem, 
we employ three waves characterized by varying speeds: the smallest denoted as $S_{l}$, 
the middle as $S_{\ast}$, and the largest as $S_{r}$. 
It is important to note that the middle wave distinguishes between the two intermediate states, 
defined by $(\rho_l^{\ast}, u_l^{\ast},p_l^{\ast})$ and $(\rho_r^{\ast}, u_r^{\ast},p_r^{\ast})$. 
For scenarios involving compressible flows, we opt for the HLLC Riemann solver \cite{toro1994restoration,toro2019hllc} 
due to its effectiveness in capturing shock discontinuities. The estimation of the smallest and largest wave speeds are as follows:
\begin{equation}\label{sl and sr}
S_{l}=\Tilde{u}-\Tilde{c}, S_{r}=\Tilde{u}+\Tilde{c},
\end{equation}
where $\Tilde{u}$ and $\Tilde{c}$ denoting the density-average particle and sound speed replacing the Roe-average variables 
in Refs. \cite{toro1994restoration,toro2019hllc} can be expressed as 
\begin{equation}
\left\{\begin{array}{l}
H_l=\frac{E_l+p_l}{\rho _l} , H_r=\frac{E_r+p_r}{\rho _r},  \Tilde{H}=\frac{\rho _l H_l+\rho _r H_r}{\rho _l+\rho _r}\\
\Tilde{u}=\frac{\rho _l u_l+\rho _r u_r}{\rho _l+\rho _r},
\Tilde{c}=\sqrt{(\gamma -1)(\Tilde{H}-0.5\Tilde{u}^2)}
\end{array},\right.
\end{equation}
with $H$ representing the enthalpy. 
The middle wave speed can be further obtained by 
\begin{equation}\label{eq:u*}
	S_{\ast}=\frac{\rho_{r} u_{r}\left(S_{r}-u_{r}\right)+\rho_{l} u_{l}\left(u_{l}-S_{l}\right)+p_{l}-p_{r}}{\rho_{r}\left(S_{r}-u_{r}\right)+\rho_{l}\left(u_{l}-S_{l}\right)}.
\end{equation}
Subsequently, we can deduce the remaining states within the star region as follows:
\begin{equation}\label{eq:p*}
p^{*}=p_{l}+\rho_{l}\left(u_{l}-S_{l}\right)\left(u_{l}-u^{*}\right)=p_{r}+\rho_{r}\left(S_{r}-u_{r}\right)\left(u^{*}-u_{r}\right),
\end{equation}
\begin{equation}\label{eq:v*}
	\mathbf{v}_{l/r}^{*}=u^{*}\mathbf{e_{ij}}+ \left[\frac{1}{2}(\mathbf{v}_{l}+\mathbf{v}_{r})- \frac{1}{2}({u}_{l}+{u}_{r})\mathbf{e_{ij}}\right],
\end{equation}
\begin{equation}\label{eq:rho*}
\rho_{l/r}^{*}=\rho_{l/r} \frac{\left(S_{l/r}-q_{l/r}\right)}{\left(S_{l/r}-u^{*}\right)},
\end{equation}
\begin{equation}\label{eq:E*}
E_{l/r}^{*}=\frac{\left(S_{l/r}-q_{l/r}\right) E_{l/r}-p_{l/r} q_{l/r}+p^{*} u^{*}}{S_{l/r}-u^{*}}.
\end{equation}
Here, $u^{*}$ is equal to $S_{\ast}$, and $q = u n_{x}+v n_{y}$, where $n_{x}$ and $n_{y}$ are components of the unit normal vector $\boldsymbol{n}$.
For weakly compressible fluid flows, assuming that the intermediate states satisfy ${p}^{*}_{l}={p}^{*}_{r}={p}^{*}$ and ${u}^{*}_{l}={u}^{*}_{r}={u}^{*}$, 
we can derive a linearized Riemann solver \cite{toro2013riemann} expressed by the following equations:
\begin{equation}\label{linearised Riemann solver}
\left\{\begin{array}{l}
u^{*}=\frac{u_{l}+u_{r}}{2}+\frac{1}{2} \frac{\left(p_{l}-p_{r}\right)}{\bar{\rho} \bar{c}} \\
p^{*}=\frac{p_{l}+p_{r}}{2}+\frac{1}{2} \bar{\rho}  \bar{c} \left(u_{l}-u_{r}\right)
\end{array}.\right.
\end{equation}
Here, $\bar{\rho}$ and $\bar{c}$ represent averages.

To decrease the numerical dissipation, 
the limiter in the HLLC Riemann solver for compressible flows is introduced \cite{wang2023extended,wang2023fourth} 
and the middle wave speed $S_{\ast}$ and the pressure in the star region can be given by 
\begin{equation}
\left\{\begin{array}{l}
u^{*}=\frac{\rho_{l}u_{l}c_{l}+\rho_{r}u_{r}c_{r}}{\rho_{l}c_{l}+\rho_{r}c_{r}}+\frac{p_{l}-p_{r}}{\rho_{l}c_{l}+\rho_{r}c_{r}}\beta^{2}_{HLLC}\\
p^{*}=\frac{p_{l}+p_{r}}{2} +\frac{1}{2}\beta_{HLLC}\left[\rho_{r}c_{r}\left(u^{*}-u_{r}\right)-\rho_{l}c_{l}\left(u_{l}-u^{*}\right)\right] 
\end{array}.\right.
\end{equation}
Here, the limiter can be derived by 
\begin{equation}\label{HLLC limiter}
\beta_{HLLC}=\min \left(\upeta_{HLLC} \left| \frac{u_{l}-u_{r}}{\bar{c}} \right|, 1\right).
\end{equation}
with $\upeta_{HLLC}=1$.

As for the weakly compressible flows, 
the middle wave speed and pressure in the linearised Riemann solver can be calculated as 
\begin{equation}\label{acoustic Riemann solver}
\left\{\begin{array}{l}
u^{*}=\frac{u_{l}+u_{r}}{2}+\frac{1}{2} \frac{\left(p_{L}-p_{R}\right)}{\bar{\rho} \bar{c}}\beta^{2}_{linearisd} \\
p^{*}=\frac{p_{l}+p_{r}}{2}+\frac{1}{2} \beta_{linearisd} \bar{\rho}  \bar{c} \left(u_{l}-u_{r}\right)
\end{array},\right.
\end{equation}
where the limiter can be given by 
\begin{equation}\label{acoustic limiter}
\beta_{linearisd}=\min \left(\upeta_{linearisd} \max (\frac{u_{l}-u_{r}}{\bar{c}}, 0), 1\right),
\end{equation}
with $\upeta_{linearisd}=15$.
%
%
\subsection{Particle relaxation technique}\label{Eulerian SPH extensions}
In practical phenomena with complex geometries, 
as the particle distribution with lattice is not sufficient, 
we implement particle relaxation \cite{zhu2021cad} to make the particles fit precisely on the surface of the complex geometry.

The geometry is imported before particle relaxation and the initial particles with lattice distribution are physically driven by 
a constant background pressure written as
\begin{equation}\label{relaxation acceleration}
    \mathbf{a}_{i}=-\frac{2}{\rho_{i}} \sum_{j}V_{j} \nabla W_{i j}.
\end{equation}
When the acceleration reaches zero, the particles not in the boundaries attain stable positions and fulfill the condition of zero-order consistency, 
which can be expressed as $\sum_{j} V_{j} \nabla W_{i j}=\mathbf{0}$. 
It is noteworthy that particles located at the boundaries do not meet the zero-order consistency criteria. 
The resolution to this issue, specifically the implementation of eulerian boundary algorithm to rectify it, 
will be elaborated upon in the forthcoming section \ref{Eulerian general boundary conditions}.
%
%
\section{General boundary algorithm within Eulerian SPH}\label{Eulerian general boundary conditions}
In this section, we elucidate the comprehensive general boundary algorithm within Eulerian SPH method through a comparative analysis 
with the boundary properties found in mesh-based FVM. 

Following Refs. \cite{wang2023extended,neuhauser2014development}, 
the discretized conservation Eq. \eqref{eqs:conservation-discretize} in both methods can be expressed uniformly as follows:
\begin{figure}
    \centering
    \includegraphics[width=1.0\textwidth]{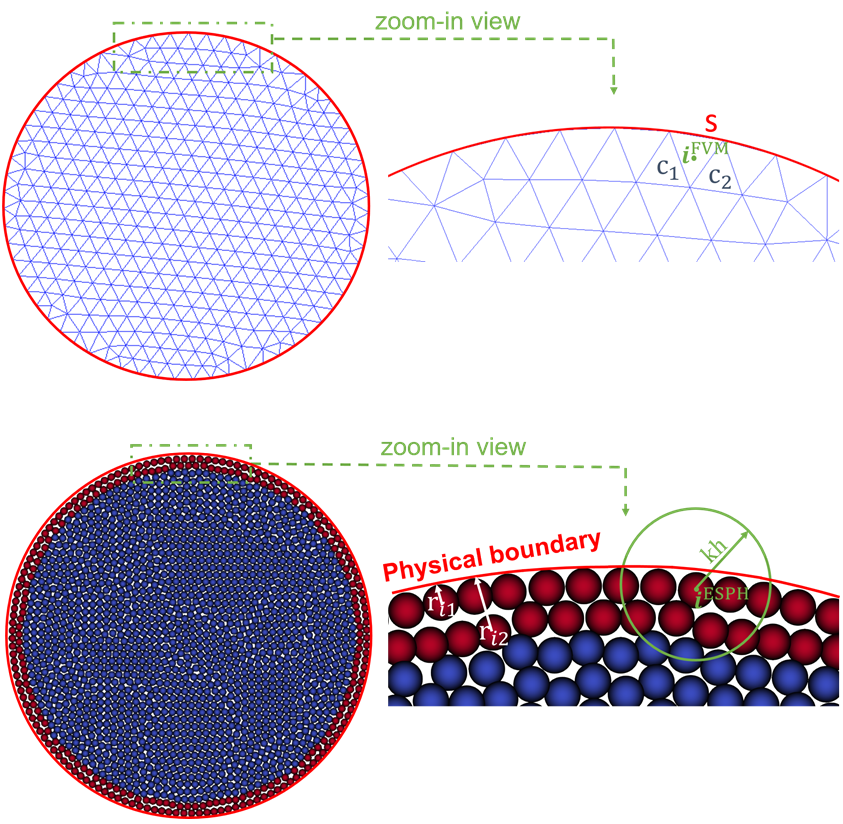}
    \caption{Mesh (top panel) and particle (bottom panel) distributions and their zoom-in views in FVM and Eulerian SPH, respectively, within a circular geometry.
    Note that the red circular line is the surface of the geometry.}
    \label{Mesh and particle distribution}
\end{figure}
\begin{equation}\label{sph and FVM common discretization}
	\frac{\partial}{\partial t}\left(\omega_{i} \mathbf{U}_{i}\right)+ \sum_{j} \mathbf{F}_{i j}\left(\mathbf{U}\right) \cdot \mathbf{A}_{i j}=0,
\end{equation}
with $j$ the neighboring cell or particle of $i$. 
Here, 
$\mathbf{A}_{i j}$ denotes the size of the interface multiplying the normal unit vector to the interface 
and determines the flux contribution from cell or particle $j$ among the total flux contributed by all neighboring cells or particles.  
Here, 
we categorize all cells in FVM and particles in Eulerian SPH into two groups: 
those within the boundaries and those outside the boundaries.
Particularly, in FVM  or in the following proposed boundary algorithm within Eulerian SPH method, 
for cells or particles located in the boundaries, 
we further subdivide the flux contribution from all cells and particles, denoted as label $j$, into two components: 
the contribution from neighboring particles or cells, denoted as label $C$ in FVM and label $P$ in Eulerian SPH method, 
within the interior in the subsequent discussion, 
and the contribution originating from the boundary surface, denoted as label $S$ in the ensuing content.

In FVM, 
cells that do not intersect with the boundaries form self-contained control volumes \cite{wang2023extended,neuhauser2014development},  
which interact exclusively with their neighboring cells, 
inherently adhering to the condition $\sum_{j} \mathbf{A}^{FVM}_{ij}=\mathbf{0}$, i.e. naturally obeying the zero-order consistency. 
However, for cells situated along the boundary surface, 
such as cell $i^{FVM}$ in Figure \ref{Mesh and particle distribution}, 
a different approach is required. 
Here, the control volume must be closed off by the boundary surface shown by the red line in Figure \ref{Mesh and particle distribution} (top panel), 
with the size of this boundary surface determining the extent of flux contribution through it. 
With this boundary surface distribution, 
the cells within the boundaries continue to satisfy the zero-order consistency requirement.
To clarify, 
the condition $\sum_{j} \mathbf{A}^{FVM}_{ij}=\mathbf{0}$ of the cells within the boundaries can be rewritten as 
\begin{equation}\label{FVM boundary consistency}
	\sum_{j} \mathbf{A}^{FVM}_{i j}=\sum_{C} \mathbf{A}^{FVM}_{i C}+ \mathbf{A}^{FVM}_{i S}=\mathbf{0}.
\end{equation}
Then, it is straightforward to derive the relation expressed as 
\begin{equation}\label{boundary consistency equation in FVM}
	\mathbf{A}^{FVM}_{i S}=-\sum_{C} \mathbf{A}^{FVM}_{i C}.
\end{equation}

In Eulerian SPH, 
the SPH approximation relies on identifying neighboring particles within the compact support \cite{gingold1977smoothed}. 
Once the particle relaxation process in Eq. \eqref{relaxation acceleration} is completed, 
particles not within the boundaries conform to the zero-order consistency principle.
However, 
for particles like particle $i^{ESPH}$ located within the boundaries, 
as clearly depicted in Figure \ref{Mesh and particle distribution} (bottom panel), 
they lack neighboring particles and thus fail to meet the zero-order consistency criteria. 
To address this issue, 
similar to Eq. \eqref{boundary consistency equation in FVM}, 
we introduce a surface vector along the normal direction to the boundary surface, which can be defined as follows:
\begin{equation}
	\mathbf{A}^{ESPH}_{i S}=-\sum_{P} \mathbf{A}^{ESPH}_{i P},
\end{equation}
with $\sum_{p} \mathbf{A}^{ESPH}_{i p}=\sum_{p}V_{p} \nabla W_{i p}$. 
Following the inclusion of contributions from the boundary surface, 
particles within the boundaries adhere to the subsequent relationship:
\begin{equation}\label{boundary consistency equation in SPH}
	\sum_{j} \mathbf{A}^{ESPH}_{i j}=\mathbf{A}^{ESPH}_{i S}+\sum_{P} \mathbf{A}^{ESPH}_{i P}=\mathbf{0}.
\end{equation}
Here, $j$ encompasses both neighboring particles and the boundary surface. 
Consequently, all particles within the computational domain achieve compliance with the zero-order consistency.

Now that we possess the algorithm for enforcing zero-order consistency in Eulerian SPH method, 
we proceed with the following steps for its implementation.
The initial step involves distinguishing whether the particles reside within the boundaries. 
As indicated in Ref. \cite{lee2008comparisons,zhang2023lagrangian}, 
we introduce the concept of the divergence of position, denoted as:
\begin{equation}\label{divergence of position}
	\nabla \cdot \mathbf{r}_{i} =\sum_{j} V_{j}\mathbf{r}_{ij} \cdot \nabla W_{i j}.
\end{equation}
Here, 
particles within the full kernel support region satisfy $\nabla \cdot \mathbf{r}_{i} \approx d$ with $d$ denoting the dimension. 
In this context, 
we select a threshold value, denoted as $\gamma$, which is set at $0.75d$ \cite{zhang2023lagrangian}. 
Consequently, 
a particle is categorized as residing within the boundaries when its divergence of position falls below this threshold value, 
and conversely, a particle is considered not to be within the boundaries.
Based on this criterion, 
we identify the particles in the outer two layers that represented by the red particle layer in Figure \ref{Mesh and particle distribution} (bottom panel) 
as being within the boundaries.
It is important to note that in FVM, 
the boundary area corresponds to the outer surface of the geometry, 
as depicted in the top panel of Figure \ref{Mesh and particle distribution}, 
which is analogous to the boundary area defined by the outermost two layers of particles in the geometry for Eulerian SPH method.
\begin{figure}
    \centering
    \includegraphics[width=1.0\textwidth]{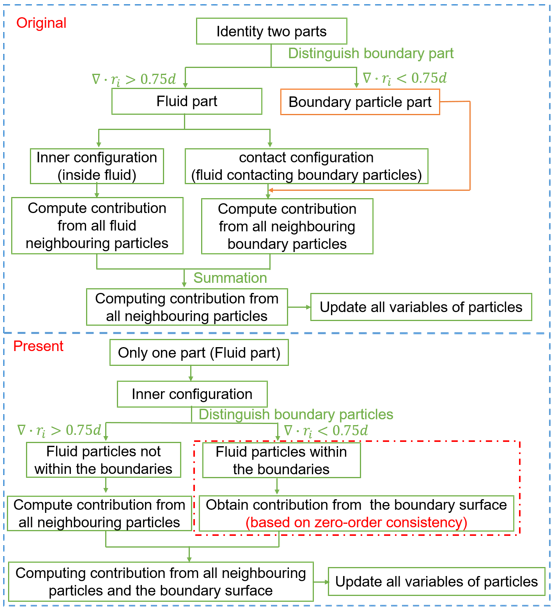}
    \caption{Flowcharts of the configurations in Eulerian SPH with the original algorithm labeled as "Original" in the top panel, 
    and the proposed algorithm labeled as "Present" in the bottom panel are presented in the case of fluid flows with designed boundary condition. 
    These flowcharts depict the process of computing contributions and updating variables at the new time based on the calculated contributions.}
    \label{flowcharts}
\end{figure}

The second step is the computation of flux contribution originating from the boundary surface $\mathbf{A}^{ESPH}_{i S}$ of each particle within the boundaries. 
Figure \ref{flowcharts} provides flowchart diagrams detailing the configurations and the computation of flux contributions for variable updates 
using the original \cite{wang2023fourth,wang2023extended,wang2023eulerian} and the proposed algorithms in Eulerian SPH 
in the case of fluid flows featuring varying boundary conditions.
Diverging from the original approach, 
where two configurations, specifically the inner and contact configurations, 
are requisite for calculating flux contributions from both fluid and boundary particle components, 
the present algorithm demands only the inner configuration.  
Following the execution of the initial step,  
the particles within the boundaries can be appropriately labeled 
and the flux contributions of these boundary particles from their neighboring particles can be accurately calculated. 
Subsequently, 
in accordance with Eq. \eqref{boundary consistency equation in SPH}, 
the flux contribution arising from the boundary surface $\mathbf{A}^{ESPH}_{i S}$ , 
as highlighted within the red dashed square in Figure \ref{flowcharts}, 
can be further obtained.
Besides, 
the distance between these particles and the physical surface of the geometry  
denoted as $r_{iS}$ similar with $r_{i1}$ and $r_{i2}$ in Figure \ref{Mesh and particle distribution} (bottom panel), 
can also be obtained easily in SPHinXsys.

The third step introduces an extra data storage space for the particles within the boundaries. 
This storage space is designed to retain essential information, 
encompassing the magnitude of the computed flux contribution from the boundary surface, i.e. $\left|\mathbf{A}^{ESPH}_{i S}\right|$, 
the normal unit vector to the boundary surface, i.e. $\frac{\sum_{p} \mathbf{A}^{ESPH}_{i p}}{\left|\sum_{p} \mathbf{A}^{ESPH}_{i p}\right|}$, 
the distance between the particle and the physical boundary surface, i.e. $r_{iS}$, 
and the input values of different boundary conditions.
In the proposed algorithm, as shown in the bottom panel of Figure \ref{flowcharts}, 
the inner configuration is divided into two parts including the inner neighbouring particles searched 
and the another configuration connecting with the boundary surface.
In SPHinXsys, 
between a pairwise particle $i$ and $j$, 
the kernel gradient $\nabla W_{i j}$ multiplying the neighbouring particle volume, 
that is the flux contribution from the particle $j$ to particle $i$,  
stored separately as the magnitude as $\frac{\partial W_{i j}}{\partial r_{ij}}V_j$ and the displacement unit vector as $\mathbf{e}_{ij}$.
Also, 
the displacement $\mathbf{r}_{ij}=\left|\mathbf{r}_{ij}\right|\mathbf{e}_{ij}$ is stored as the magnitude $\left|\mathbf{r}_{ij}\right|$ 
and replacement unit vector $\mathbf{e}_{ij}$.
Based on this data structure, 
the flux contribution from the boundary surface of $-\left|\sum_{p} \mathbf{A}^{ESPH}_{i p}\right|$ is stored in 
data storage of $\frac{\partial W_{i j}}{\partial r_{ij}}V_j$ and  
normal unit vector to the boundary surface of $-\frac{\sum_{p} \mathbf{A}^{ESPH}_{i p}}{\left|\sum_{p} \mathbf{A}^{ESPH}_{i p}\right|}$ is stored as $\mathbf{e}_{ij}$.
Besides, 
the distance between the boundary particles and the physical surface of $r_{iS}$ is stored in the data structure of $\left|\mathbf{r}_{ij}\right|$. 

The fourth step entails the implementation of various boundary conditions. 
This algorithm exhibits a high level of generality, making it suitable for accommodating a wide array of boundary conditions 
such as the reflecting, inflow, outflow, motiontive, non-reflecting, 
non-slip wall and far-field boundaries. 
For each boundary condition, 
we input the designated values of the boundary surface in the storage space. 
Since we have obtained the flux contributions from the the boundary surface separately in the third step, 
we can readily calculate the flux from the boundary surface with the given boundary condition. 
Subsequently, 
the total flux, encompassing contributions from neighboring particles and the boundary surface, can be obtained, 
and this information is used to update the variables for the new timestep.
It is worth noting that in the original algorithm the boundary conditions were computed by relying on the many boundary particles in the contact configuration, 
whereas in the present algorithm the boundary conditions are computed by relying on the values of the extra storage space 
in the inner configuration rather than the particles.
Therefore, this boundary algorithm has the added benefit of reducing some computational effort as less neighbouring particles needed.
%
%
\section{Numerical results}\label{Numerical results}
In this section, 
several numerical examples including fully and weakly compressible flows with the complicated geometries 
are studied to validate the stability and accuracy of the proposed boundary algorithm using different boundary conditions 
in Eulerian SPH method.
In all cases, 
the truncated Laguerre-Gauss kernel \cite{wang2023fourth} is applied for high accuracy with the cut-off radius as $2.6dp$ 
where $dp$ is the initial particle spacing.
%
%
\subsection{A Mach 3 wind tunnel with a step}
\begin{figure}
    \centering
    \includegraphics[width=1.0\textwidth]{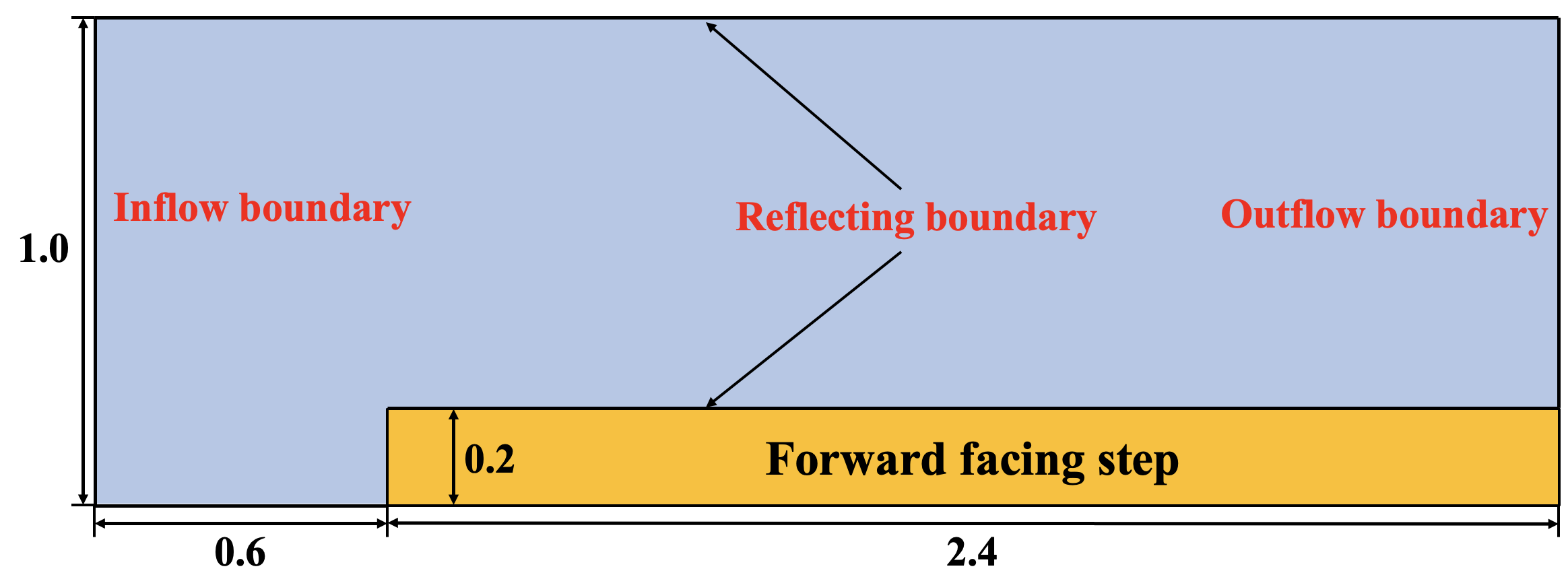}
    \caption{A Mach $3$ wind tunnel with a step: The geometry and boundary conditions.}
    \label{wind tunnel with a step: The geometry and boundary conditions}
\end{figure}

In this section, we delve into the examination of a Mach 3 inviscid flow around a two-dimensional forward-facing step. 
The primary objective is to validate the effectiveness of the proposed algorithm for handling boundary conditions, 
encompassing inflow, outflow, and reflective conditions, 
within this specific scenario.
Following Refs. \cite{woodward1984numerical,wen2023space,zhu2017new,wang2016developing}, 
the wind tunnel in question boasts dimensions of one unit in width and three units in length. 
Positioned within this tunnel is a step, 
which is situated at a distance of 0.6 units from the tunnel's left-hand extremity and exhibits a height of 0.2 units.
The left is the inflow boundary condition and the right is the supersonic outflow boundary condition with no effect on the flow, 
as well as the top and bottom are reflecting boundary conditions as is shown in Figure \ref{wind tunnel with a step: The geometry and boundary conditions}
which provides the geometry and boudnary conditions.
\begin{figure}
    \centering
    \includegraphics[width=1.0\textwidth]{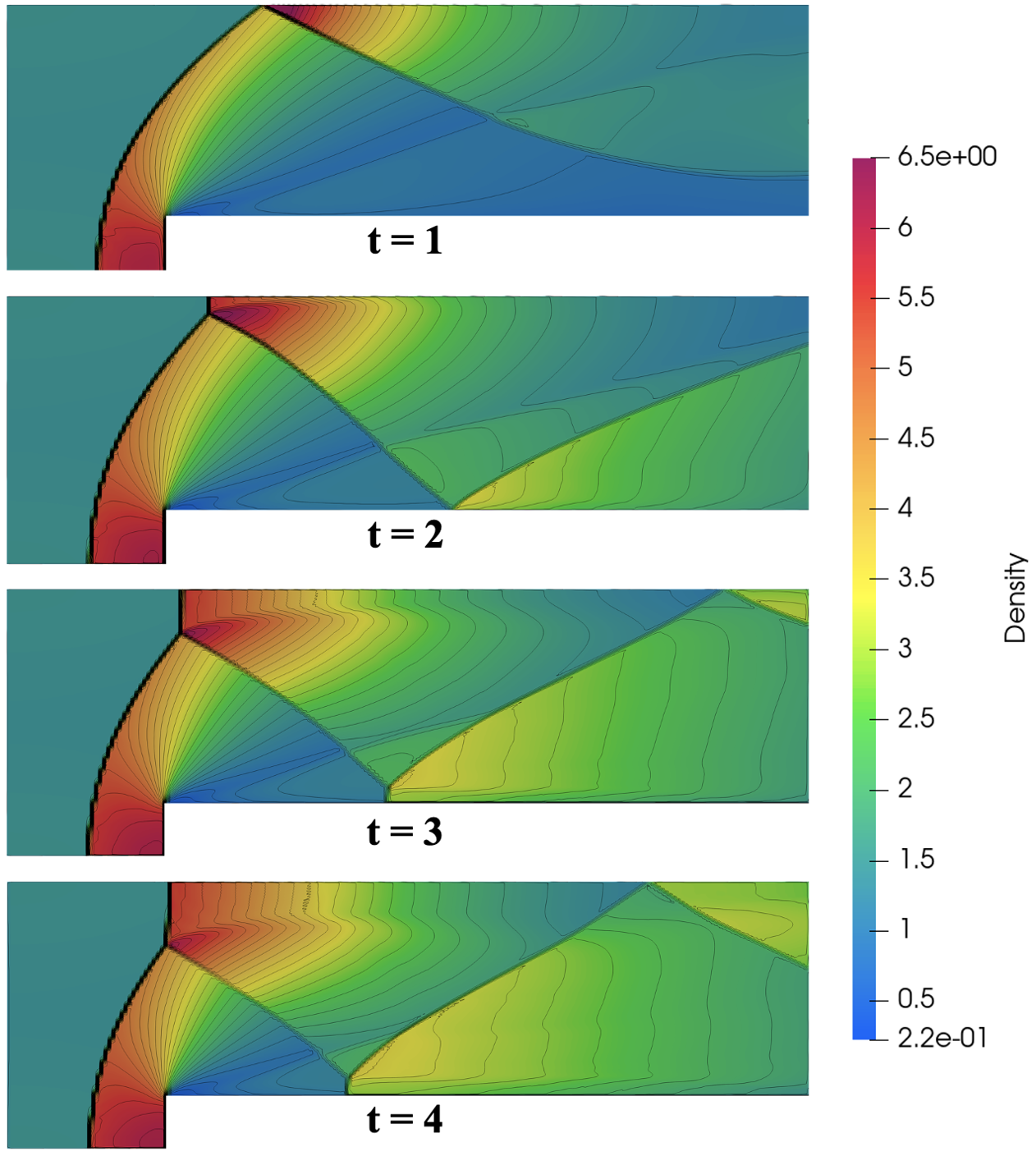}
    \caption{A Mach $3$ wind tunnel with a step: 
    $30$ equally spaced density contour ranging from $0.22$ to $6.5$ with the spatial resolution $dp=1/200$ at the different instants 
    including $t=1$ (top panel), $t=2$ (second top panel), $t=3$ (third top panel) and $t=4$ (bottom panel).}
    \label{wind tunnel with a step: density contour}
\end{figure}

Figure \ref{wind tunnel with a step: density contour} illustrates the density contours over a range of $0.22$ to $6.5$, 
achieved using a spatial resolution of $dp=1/200$ at various time steps from $t=1$ to $t=4$. 
This figure demonstrates the remarkable smoothness of the results obtained through the proposed boundary algorithm, 
thereby affirming the algorithm's robustness.
Moreover, 
it is worth noting that the density contours produced with this boundary algorithm closely align with findings from prior works, 
specifically those referenced in Refs. \cite{woodward1984numerical,wen2023space,zhu2017new,wang2016developing}, 
validating the accuracy and convincing of the proposed algorithm.
%
%
\subsection{Double Mach reflection of a strong shock}
\begin{figure}
    \centering
    \includegraphics[width=1.0\textwidth]{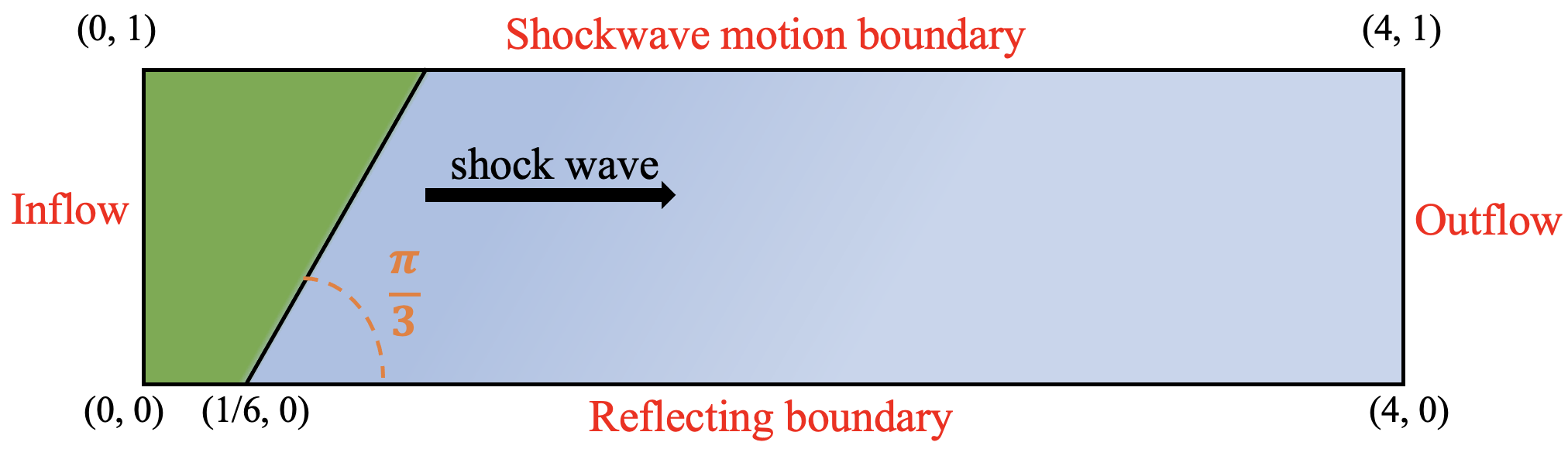}
    \caption{Double Mach reflection of a strong shock: The geometry and boundary conditions.}
    \label{Double Mach reflection of a strong shock: The geometry and boundary conditions}
\end{figure}

In this section, 
a two-dimensional inviscid flow, i.e. double Mach reflection of a strong shock, 
is tested to investigate the stability of the boundary algorithm with more complex boundary boundaries. 
The boundaries in this case contained various types including inflow, outflow, shock wave motion, 
and reflecting boundary conditions shown in Figure \ref{Double Mach reflection of a strong shock: The geometry and boundary conditions}.
Following Ref. \cite{woodward1984numerical}, 
the conducted test pertains to a Mach 10 shock wave propagating through the air with a specific heat ratio $\gamma$ of $1.4$, 
and this shock wave forms a 60-degree angle with a reflective wall initially.
Besides, 
the computational domain encompasses the region defined by $(x,y) \in [0,4]\times[0,1]$, 
and the initial condition is specified as follows:
\begin{equation}
	(\rho,u,v,p)= \begin{cases}(1.4,0,0,1) &  y\leq1.732(x-0.1667) \\ (8,7.145,-4.125,116.8333) & \text { otherwise }\end{cases},
\end{equation}
with the final time as $t = 0.2$ and the resolution is $dp=1/120$ to discretize the computational domain.
Specifically, 
the bottom boundary is defined as a reflecting wall, initiating at $x=1/6$ where the shock is inclined at a 60-degree angle with respect to the x-axis. 
It extends vertically to the top of the domain at $y=1$. The initial post-shock input boundary condition is applied from $x=0$ to $x=1/6$. 
The left-hand boundary is set as the assigned value as the initial post-shock boundary condition, 
while the right-hand boundary is set to have zero-gradient output conditions.
Moreover, 
for the top boundary, 
the variable is introduced to accurately describe the motion of the initial Mach $10$ shock.

\begin{figure}
    \centering
    \includegraphics[width=1.0\textwidth]{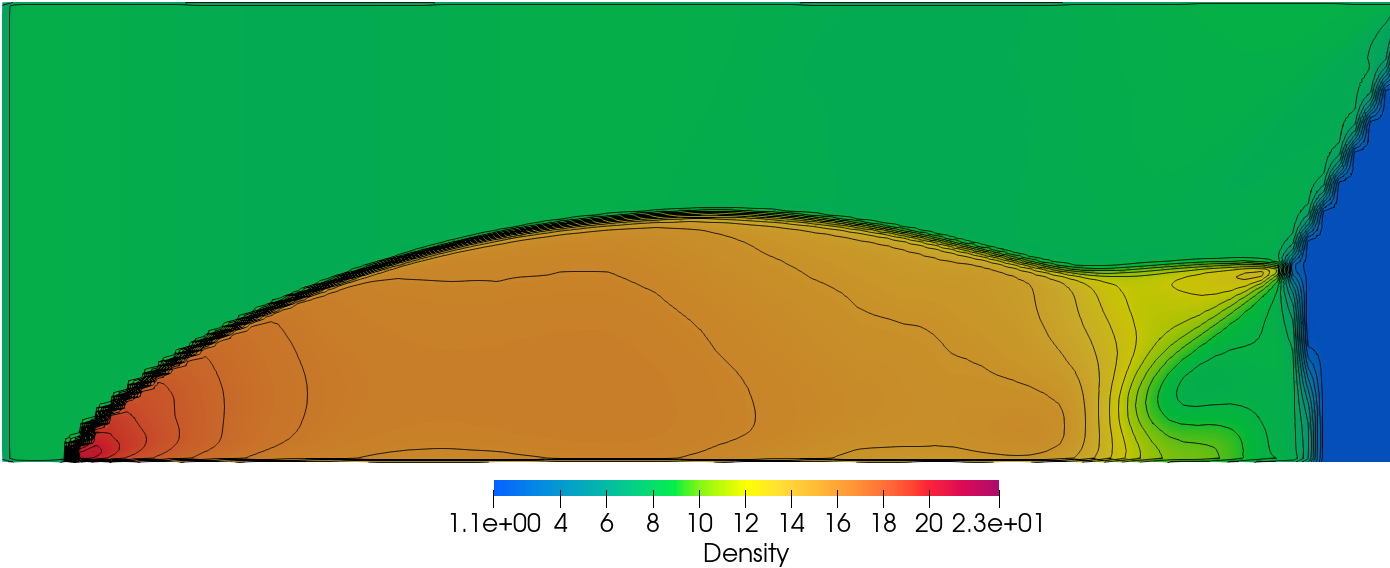}
    \caption{Double Mach reflection of a strong shock: 
    $30$ equally spaced density contour ranging from $1.1$ to $23.0$ with the spatial resolution $dp=1/120$ at the final time $t=0.2$.}
    \label{Double Mach reflection of a strong shock}
\end{figure}
Figure \ref{Double Mach reflection of a strong shock} presents the density contour 
ranging from $1.1$ to $23.0$ under the resolution as $dp=1/120$ at the final time $t=0.2$.
It can be observed that the density contour is smooth using the different and complex boundary conditions 
and agrees well with the results in Refs. \cite{woodward1984numerical,wang2023extended}, 
validating that the proposed boundary algorithm is stable and accurate.
%
%
\subsection{Supersonic flow past a circular cylinder}
\begin{figure}
    \centering
    \includegraphics[width=0.6\textwidth]{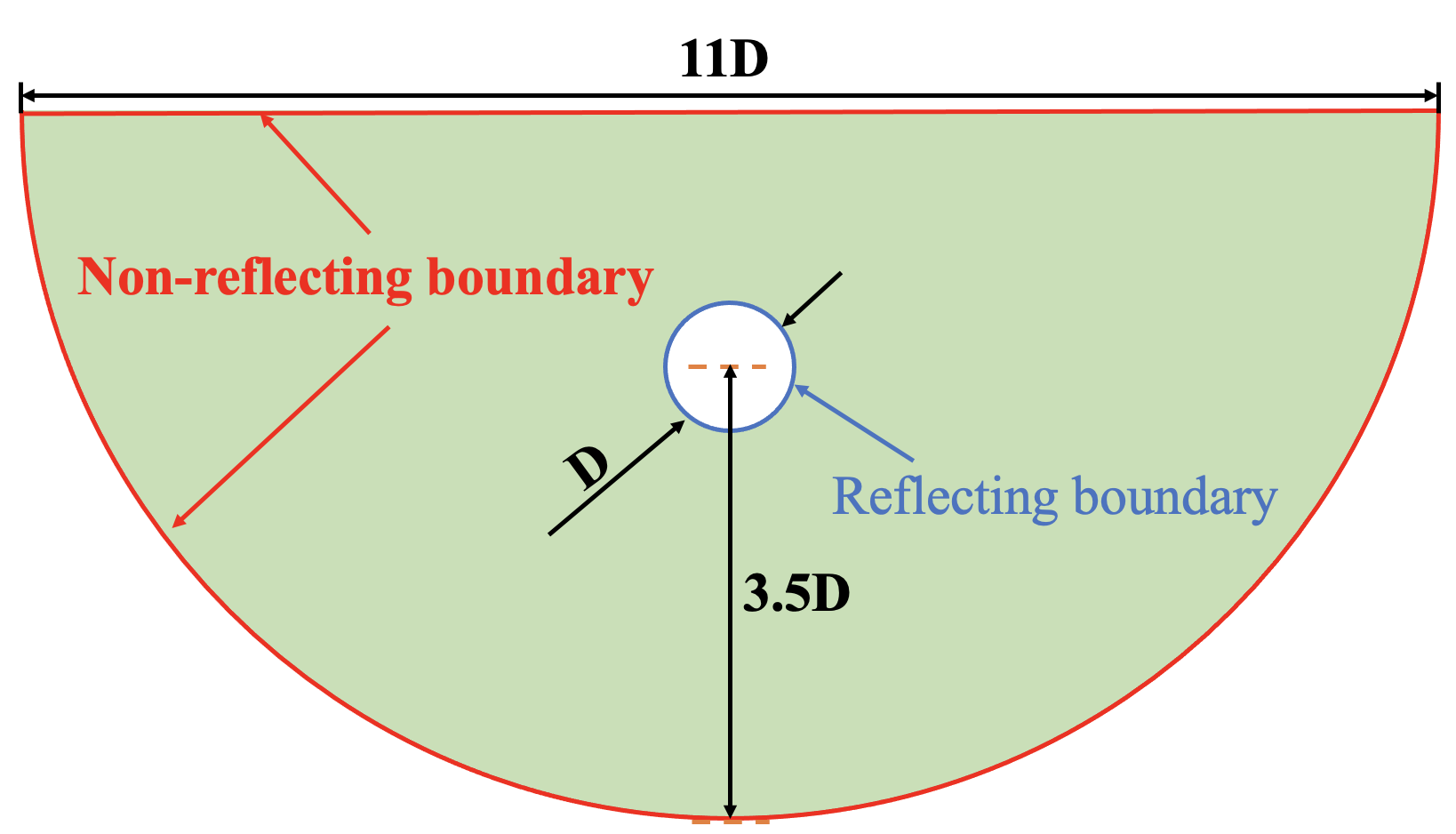}
    \caption{Supersonic flow past a circular cylinder: The geometry and boundary conditions.}
    \label{Supersonic flow past a circular cylinder: The geometry and boundary conditions}
\end{figure}

In this section, 
we explore supersonic flow with complex geometry. 
Specifically, 
we simulate supersonic flow past a circular cylinder to assess the robustness of the proposed algorithm 
in handling intricate geometries under the different boundary conditions including non-reflecting and reflecting boundary conditions.
Following Refs. \cite{zhu2019free,sinclair2017theoretical}, 
the reflecting boundary condition is imposed on the cylinder surface, 
while the non-reflecting boundary conditions including supersonic inflow and outflow conditions are enforced at the other boundaries.
In the case, 
the pressure and density inflows are set as $\rho_{\infty}=1.0$ and  $p_{\infty}=1/\gamma$ with $\gamma=1.4$, respectively.
Besides, 
we configure the computational domain as a semi-circular geometry with a radius of $5.5D$, 
and the center of the inner cylinder is positioned at a distance of $3.5D$ from the inflow boundary 
shown as Figure \ref{Supersonic flow past a circular cylinder: The geometry and boundary conditions}, 
with a cylinder diameter of $D$ under the spatial resolution of $dp=1/20$ to discretize the computational domain.

\begin{figure}
    \centering
    \includegraphics[width=0.6\textwidth]{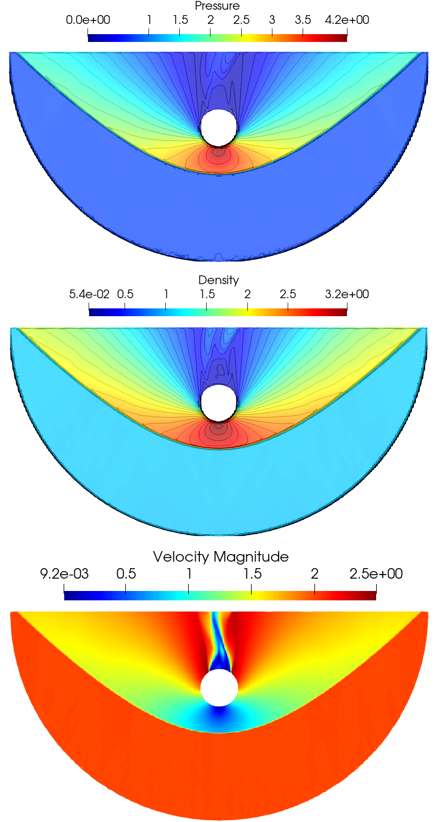}
    \caption{Supersonic flow past a circular cylinder (Mach $2$): 
    The pressure contour ranging from $0.0$ to $4.2$ (top panel) and density contour ranging from $0.054$ to $3.2$ (middle panel) 
    and velocity magnitude distribution ranging from $0.0092$ to $2.5$ (bottom panel) at the spatial resolution $dp=1/20$.}
    \label{Supersonic flow past a circular cylinder Mach 2}
\end{figure}

\begin{figure}
    \centering
    \includegraphics[width=0.6\textwidth]{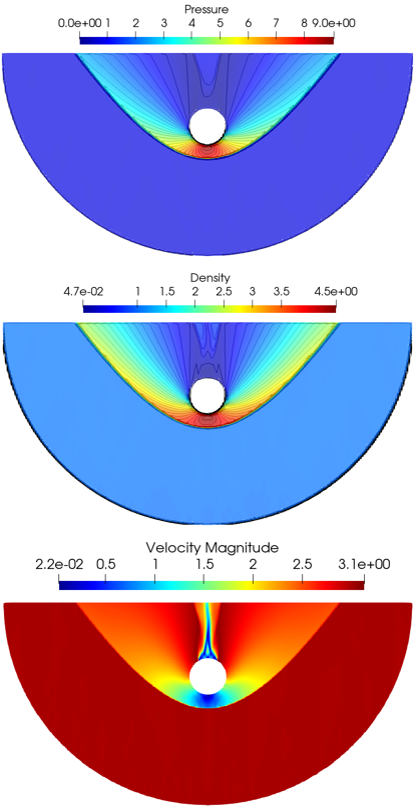}
    \caption{Supersonic flow past a circular cylinder (Mach $3$): 
    The pressure contour ranging from $0.0$ to $8.9$ (top panel) and density contour ranging from $0.047$ to $4.5$ (middle panel) 
    and velocity magnitude distribution ranging from $0.022$ to $3.1$ (bottom panel) at the spatial resolution $dp=1/20$.}
    \label{Supersonic flow past a circular cylinder Mach 3}
\end{figure}

Figures \ref{Supersonic flow past a circular cylinder Mach 2} and \ref{Supersonic flow past a circular cylinder Mach 3} show the pressure, density and velocity contours under the spatial resolution $dp=1/20$ with $\text{Mach} = 2$ and $\text{Mach} = 3$, respectively.
It can be observed that the proposed boundary algorithm for non-reflecting and reflecting boundary conditions can obtain the smooth results and enable capture the shock property, representing its stability and accuracy.
Besides, 
the results in Figures \ref{Supersonic flow past a circular cylinder Mach 2} and \ref{Supersonic flow past a circular cylinder Mach 3} agree well with the results in Refs. \cite{zhu2019free,sinclair2017theoretical} with $\text{Mach} = 2$ and $\text{Mach} = 3$, 
proving the results with non-reflecting and reflecting boundary conditions containing the complex geometry are convincing and correct.

%
%
%
\subsection{Flow around a circular cylinder}
\begin{figure}
    \centering
    \includegraphics[width=0.7\textwidth]{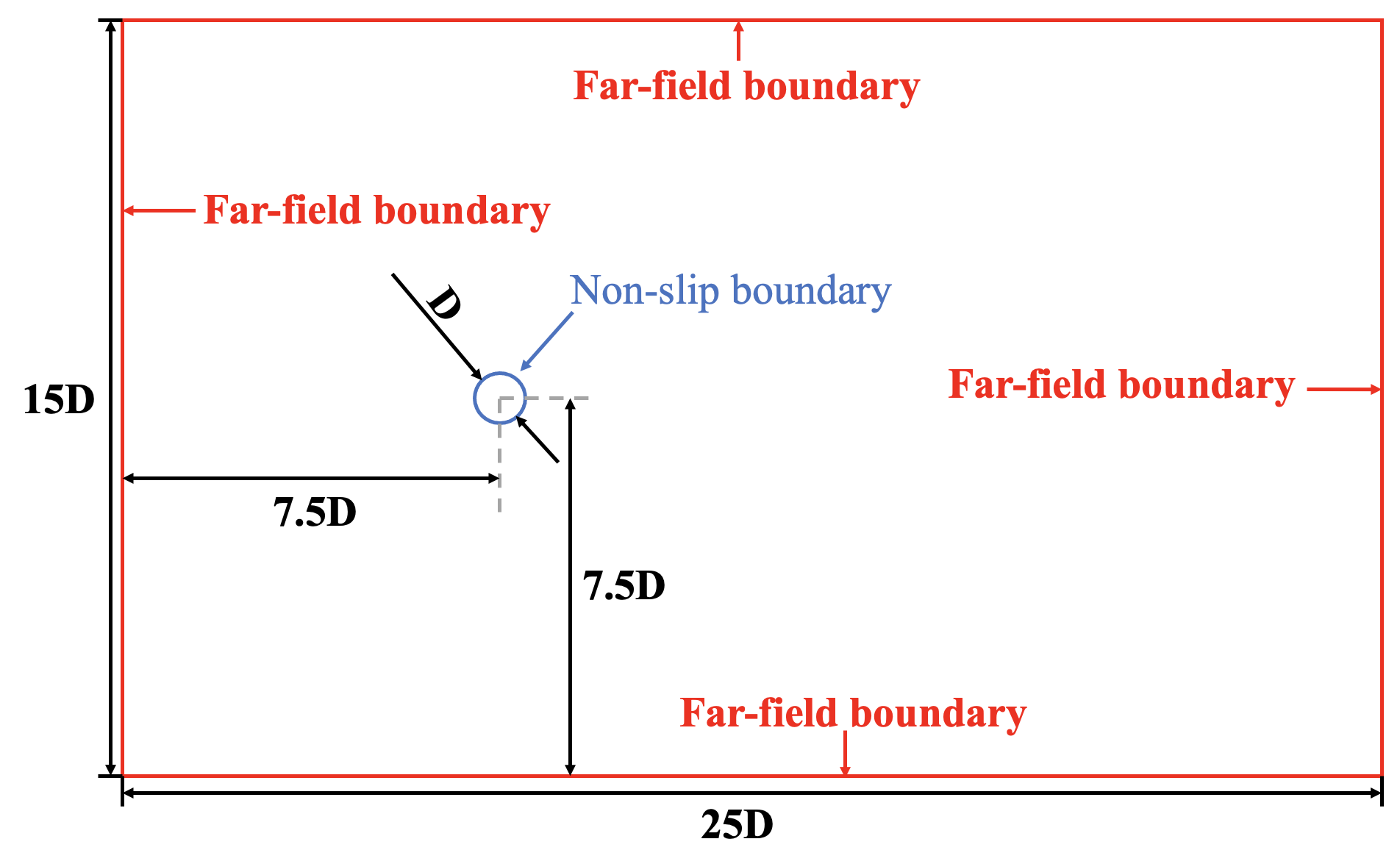}
    \caption{Flow around a circular cylinder: The geometry and boundary conditions.}
    \label{Flow past a circular cylinder: The geometry and boundary conditions}
\end{figure}

In this section, 
we investigate the impact of the proposed boundary algorithm in a two-dimensional viscous flow scenario, i.e. the flow around a circular cylinder. 
This problem features complex geometry and incorporates various boundary conditions, including non-slip wall and far-field conditions. 
As illustrated in Figure \ref{Flow past a circular cylinder: The geometry and boundary conditions}, 
which provides a depiction of the geometry and the distinct boundary conditions, 
we designate the circular surface as the non-slip wall boundary condition, 
while the remaining boundaries are configured as the far-field boundary condition.
Following the approach outlined in Ref. \cite{wang2023fourth}, 
to mitigate the influence of the far-field boundary conditions, 
a computational domain of substantial size, namely $[25D, 15D]$, is employed. 
The center of the cylinder is positioned at $(7.5D, 7.5D)$, with the cylinder itself possessing a diameter of $D=2$.
To perform a quantitative analysis of the accuracy of the results, 
we employ drag and lift coefficients, 
which are defined as:
\begin{equation}\label{eq:wavespeed}
C_{D}=\frac{2F_{D}}{\rho_{\infty}u_{\infty}^2 A},    C_{L}=\frac{2F_{L}}{\rho_{\infty}u_{\infty}^2 A}.
\end{equation}
Here, $F_{D}$ and $F_{L}$ represent the drag and lift forces acting on the cylinder, respectively.
Besides, 
a constant far-field velocity $u_{\infty}$ and density $\rho_{\infty}$ are both set to $1$.
Also, the Strouhal number $St=fD/u_\infty$ with $f$ the frequency in the unsteady cases 
and the Reynolds numbers $Re=\rho_{\infty}u_{\infty}D/\mu$ are set as $100$ and $200$ in the case and the computational time is $t=300$. 
For the convergence study, 
the spatial resolutions are applied as $dp=10$, $20$, and $30$.

\begin{figure}
    \centering
	\begin{subfigure}[b]{0.65\textwidth}
		\centering
		\includegraphics[trim = 0cm 0cm 0cm 0cm, clip, width=0.9\textwidth]{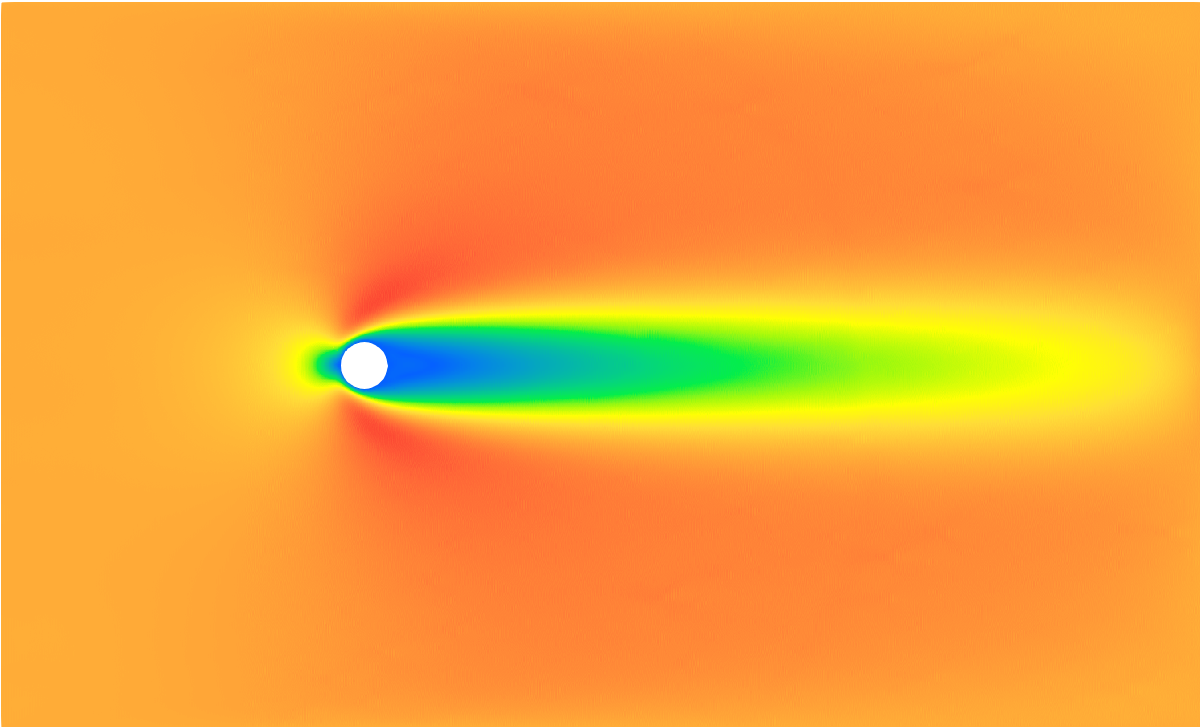}
		\caption{The Reynolds number $Re=20$.}
	\end{subfigure}
	\centering
	\begin{subfigure}[b]{0.65\textwidth}
		\centering
		\includegraphics[trim = 0cm 0cm 0cm 0cm, clip, width=0.9\textwidth]{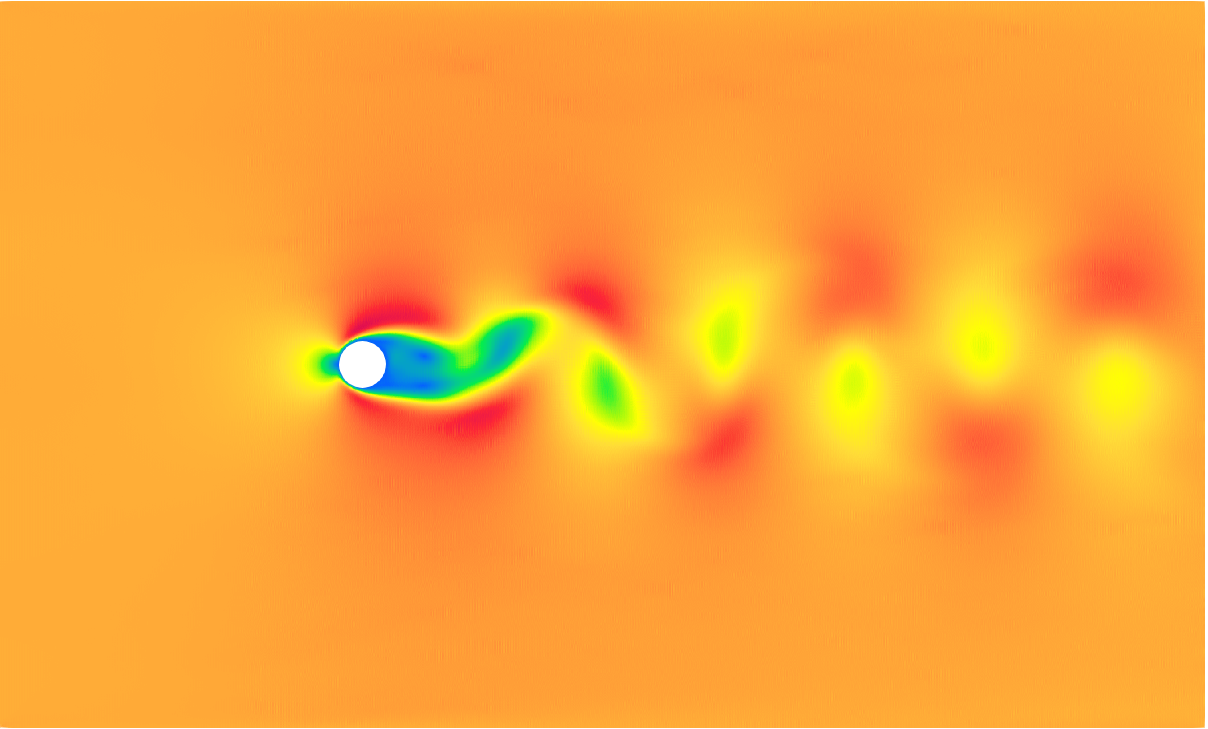}
		\caption{The Reynolds number $Re=100$.}
	\end{subfigure}
	\centering
	\begin{subfigure}[b]{0.65\textwidth}
		\centering
		\includegraphics[trim = 0cm 0cm 0cm 0cm, clip, width=0.9\textwidth]{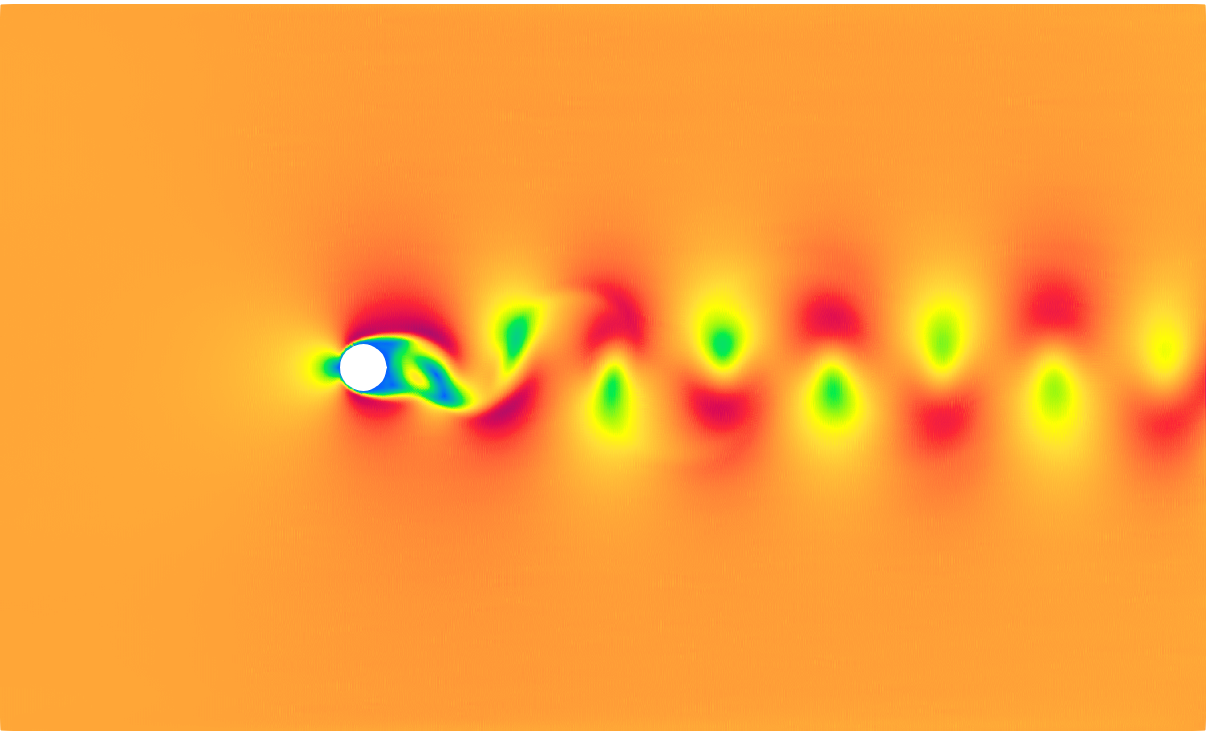}
		\caption{The Reynolds number $Re=200$.}
	\end{subfigure}
	
	\caption{Flow around a circular cylinder: The velocity distributions ranging from $6.0\times 10^{-4}$ to $1.5$ under the different Reynolds numbers at the finial time $t=300$.}
	\label{Flow around a circular cylinder: The velocity distributions}
\end{figure}
\begin{figure}
    \centering
	\begin{subfigure}[b]{0.9\textwidth}
		\centering
		\includegraphics[trim = 0cm 0cm 0cm 0cm, clip, width=0.95\textwidth]{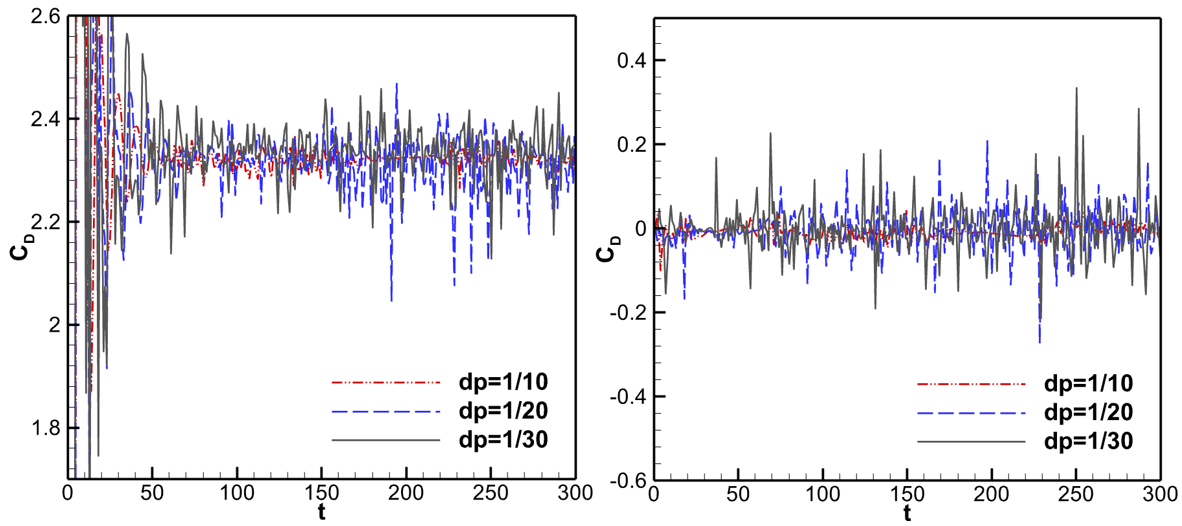}
		\caption{The Reynolds number $Re=20$.}
	\end{subfigure}
	\centering
	\begin{subfigure}[b]{0.9\textwidth}
		\centering
		\includegraphics[trim = 0cm 0cm 0cm 0cm, clip, width=0.95\textwidth]{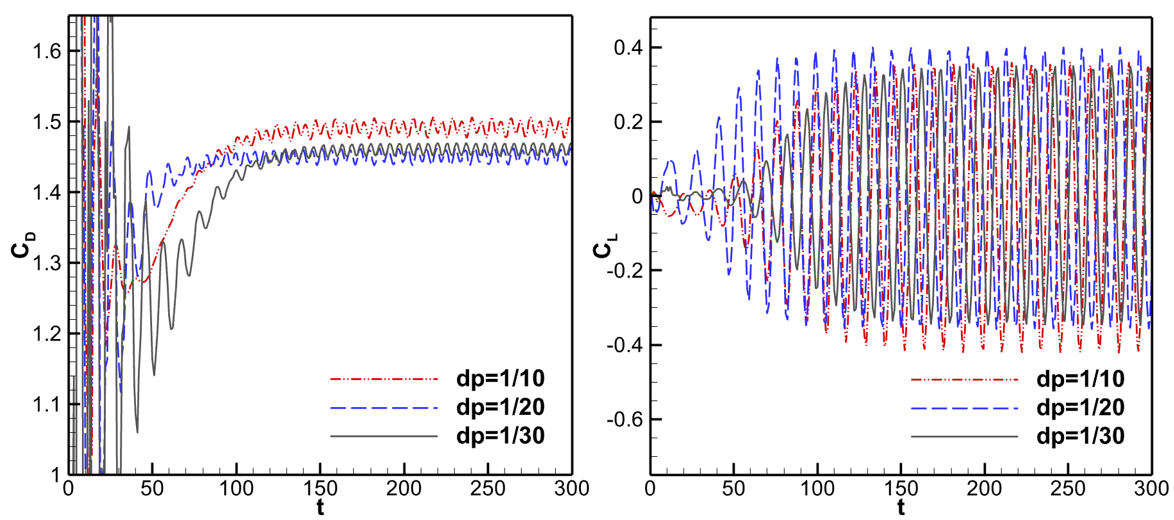}
		\caption{The Reynolds number $Re=100$.}
	\end{subfigure}
	\centering
	\begin{subfigure}[b]{0.9\textwidth}
		\centering
		\includegraphics[trim = 0cm 0cm 0cm 0cm, clip, width=0.95\textwidth]{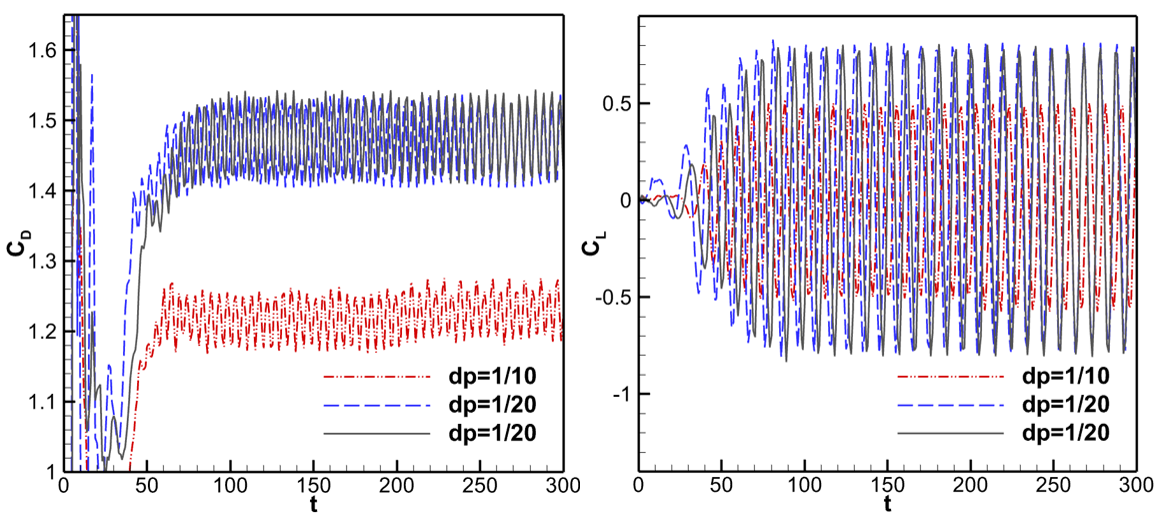}
		\caption{The Reynolds number $Re=200$.}
	\end{subfigure}
	
	\caption{Flow around a circular cylinder: The convergence study of the drag and lift coefficients $C_D$ and $C_L$ under different Reynolds numbers with the spatial resolution $dp=1/30$.}
	\label{Flow around a circular cylinder: The convergence study}
\end{figure}
\begin{table}
    \caption{Flow around a cylinder: Drag and lift coefficients from different experimental and simulation results with $Re=20$ 
	and the result using the proposed boundary algorithm with the resolution $dp=1/30$. }
    \centering
    \begin{tabular}{cc}
    \hline
    \begin{tabular}[c]{@{}c@{}} \end{tabular} &
    \begin{tabular}[c]{@{}c@{}}$C_{D}$\end{tabular}  \\ \hline
    Tritton \cite{tritton1959experiments} & 2.09 \\ \hline
    Taira and Colonius \cite{taira2007immersed} & 2.06  \\ \hline
    Tafuni et al. \cite{tafuni2018versatile} & 2.29  \\ \hline
    Negi et al. \cite{negi2020improved} & 2.32  \\ \hline
    Present & 2.32  \\ \hline
    \end{tabular}
    \label{Table_Re=20}
\end{table}
\begin{table}
    \caption{Flow around a cylinder: Drag and lift coefficients from different experimental and simulation results with $Re=100$ 
	and the result using the proposed boundary algorithm with the resolution $dp=1/30$. }
    \centering
    \begin{tabular}{cccc}
    \hline
    \begin{tabular}[c]{@{}c@{}} \end{tabular} &
    \begin{tabular}[c]{@{}c@{}}$C_{D}$\end{tabular} &
     \begin{tabular}[c]{@{}c@{}}$C_{L}$\end{tabular} &
     \begin{tabular}[c]{@{}c@{}}$S_{t}$\end{tabular} \\ \hline
    White\cite{white2006viscous} & 1.46 & -  & - \\ \hline
    Chiu et al.\cite{chiu2010differentially} & 1.35 $\pm$ 0.012 & $\pm$0.303  & 0.166  \\ \hline
    Le et al.\cite{le2006immersed} & 1.37 $\pm$ 0.009 & $\pm$0.323 & 0.160  \\ \hline
    Brehm et al.\cite{brehm2015locally} & 1.32 $\pm$ 0.010 & $\pm$0.320 & 0.165  \\ \hline
    Liu et al.\cite{liu1998preconditioned} & 1.35 $\pm$ 0.012 & $\pm$0.339 & 0.165  \\ \hline
    Zhang et al.\cite{zhang2023lagrangian} & 1.61 $\pm$ 0.005 & $\pm$0.448 & 0.171  \\ \hline
    Present & 1.46 $\pm$ 0.009 & $\pm$0.349 & 0.176  \\ \hline
    \end{tabular}
    \label{Table_Re=100}
\end{table}
\begin{table}
    \caption{Flow around a cylinder: Drag and lift coefficients from different experimental and simulation results with $Re=200$ 
	and the result using the proposed boundary algorithm with the resolution $dp=1/30$. }
    \centering
    \begin{tabular}{cccc}
    \hline
    \begin{tabular}[c]{@{}c@{}} \end{tabular} &
    \begin{tabular}[c]{@{}c@{}}$C_{D}$\end{tabular} &
     \begin{tabular}[c]{@{}c@{}}$C_{L}$\end{tabular} &
     \begin{tabular}[c]{@{}c@{}}$S_{t}$\end{tabular} \\ \hline
   Taira et al. \cite{taira2007immersed} & 1.35 $\pm$ 0.048 & $\pm$0.68  & 0.196  \\ \hline
    Tafuni et al. \cite{tafuni2018versatile} & 1.46 & $\pm$0.693  & 0.206  \\ \hline
    Negi et al. \cite{negi2020improved} & 1.524 $\pm$ 0.05 & $\pm$0.722 & 0.210  \\ \hline
    Jin and Braza \cite{jin1993nonreflecting} & 1.532 $\pm$ 0.05 & $\pm$0.744 & 0.210  \\ \hline
    Zhang et al.\cite{zhang2023lagrangian} & 1.63 $\pm$ 0.05 & $\pm$0.84 & 0.212  \\ \hline
    Present & 1.48 $\pm$ 0.055 & $\pm$0.806 & 0.208  \\ \hline
    \end{tabular}
    \label{Table_Re=200}
\end{table}

Figure \ref{Flow around a circular cylinder: The velocity distributions} 
portrays the velocity distributions ranging from $6.0\times 10^{-4}$ to $1.5$ 
with the different Reynolds numbers including $Re=20$, $100$ and $200$ at the finial time $t=300$, 
showing that smooth velocity results with different Reynolds numbers can be achieved by using the proposed algorithm. 
Also, 
the velocity distribution and the vortex street appeared during the simulation are agreement well with the results in Ref. \cite{zhang2023lagrangian}.
Besides, 
with the increase of the resolutions from $dp=1/10$ to $dp=1/30$, 
the results converge rapidly shown in Figure \ref{Flow around a circular cylinder: The convergence study}.
It is observed that these plots reveal that the drag coefficients stabilize to a consistent mean value after an initial period of fluctuation, 
while the lift coefficient oscillates around zero. 
The deviations in the drag and lift coefficients for spatial resolutions of $dp=1/20$ and $dp=1/30$ remain below 3 percent, 
and the frequencies and amplitudes of the lift coefficient are approximately consistent. 
The converged results are listed in Tables \ref{Table_Re=20}, \ref{Table_Re=100} and \ref{Table_Re=200} 
which contains a series of numerical and experimental results with the Reynolds numbers $Re=20$, $Re=100$ and $Re=200$, respectively.
The observation shows that these results are agreement with the references in above Tables, 
validating that the proposed algorithm is convincing and accurate.  
In this study, 
the computations have been executed on a desktop computer equipped with an Intel Core i7-10700 processor, 
boasting 8 cores and a clock speed of 2.90 GHz. 
The total CPU wall-clock time consumed by the original boundary algorithm \cite{wang2023fourth,wang2023extended}, 
with the spatial resolution $dp=1/10$ and Reynolds number $Re=100$ throughout the entire process, amounts to $4141.94$ seconds. 
However, 
the wall-clock time required by the proposed algorithm is $3600.63$ seconds, 
showing its higher computational efficiency.
%
%
\section{Summary and conclusion}\label{Summary and conclusions}
In this paper, 
we firstly implement the particle relaxation technique in Eulerian SPH method generating the body-fitted particle distribution 
and satisfying the zero-order consistency of the particles not within the boundaries.
By analysing the boundary algorithm in FVM, 
we propose the analogous algorithm in Eulerian SPH in detail 
and after implementation, the particles within the boundaries therefore also satisfy the zero-order consistency.
Moreover, 
several numerical cases including fully and weakly compressible flows are tested and 
the results validate the stability and accuracy of the proposed algorithm.
Another supplementary advantage lies in its potential to enhance computational efficiency to a certain degree.
%
%
\bibliographystyle{elsarticle-num}
\bibliography{ref}
\end{document}